\documentclass[sigconf]{acmart}
\AtBeginDocument{%
  }

\usepackage{microtype}
\usepackage{graphicx}
\usepackage{subfigure}
\usepackage{booktabs} 
\usepackage{hyperref}
\usepackage{extramarks}
\usepackage{enumitem}
\usepackage{fancyhdr}
\usepackage{booktabs}
\usepackage{array}
\usepackage{listings}
\usepackage{xcolor}

\usepackage{float}

\setcopyright{acmlicensed}
\copyrightyear{2025}
\acmYear{2025}
\acmDOI{3706599.3719790}
  
\newcommand{\name}{\textsc{Interactive Sketchpad}}

\copyrightyear{2025}
\acmYear{2025}
\setcopyright{rightsretained}
\acmConference[CHI EA '25]{Extended Abstracts of the CHI Conference on Human Factors in Computing Systems}{April 26-May 1, 2025}{Yokohama, Japan}
\acmBooktitle{Extended Abstracts of the CHI Conference on Human Factors in Computing Systems (CHI EA '25), April 26-May 1, 2025, Yokohama, Japan}\acmDOI{10.1145/3706599.3719790}
\acmISBN{979-8-4007-1395-8/2025/04}
\begin{document}

\author{Jimin Lee}
\authornote{Equal Contribution.}
\email{jimin24@mit.edu}
\affiliation{%
  \institution{Massachusetts Institute of Technology}
  \city{Cambridge}
  \state{MA}
  \country{USA}
}

\author{Steven-Shine Chen}
\authornotemark[1]
\email{stevensh@mit.edu}
\affiliation{%
  \institution{Massachusetts Institute of Technology}
  \city{Cambridge}
  \state{MA}
  \country{USA}
}

\author{Paul Pu Liang}
\email{ppliang@mit.edu}
\affiliation{%
  \institution{Massachusetts Institute of Technology}
  \city{Cambridge}
  \state{MA}
  \country{USA}
}

\title{\name: A Multimodal Tutoring System for Collaborative, Visual Problem-Solving}

\begin{abstract}
Humans have long relied on visual aids like sketches and diagrams to support reasoning and problem-solving. Visual tools, like auxiliary lines in geometry or graphs in calculus, are essential for understanding complex ideas. However, many tutoring systems remain text-based, providing feedback only through natural language. Leveraging recent advances in Large Multimodal Models (LMMs), this paper introduces \name, a tutoring system that combines language-based explanations with interactive visualizations to enhance learning. Built on a pre-trained LMM, \name\ is fine-tuned to provide step-by-step guidance in both text and visuals, enabling natural multimodal interaction with the student. Accurate and robust diagrams are generated by incorporating code execution into the reasoning process. User studies conducted on math problems such as geometry, calculus, and trigonometry demonstrate that \name\ leads to improved task comprehension, problem-solving accuracy, and engagement levels, highlighting its potential for transforming educational technologies. All code is available at: \url{https://stevenshinechen.github.io/interactivesketchpad/}.
\end{abstract}

\begin{teaserfigure}
    \centering
    \vspace{-2mm}
    \includegraphics[width=0.8\textwidth]{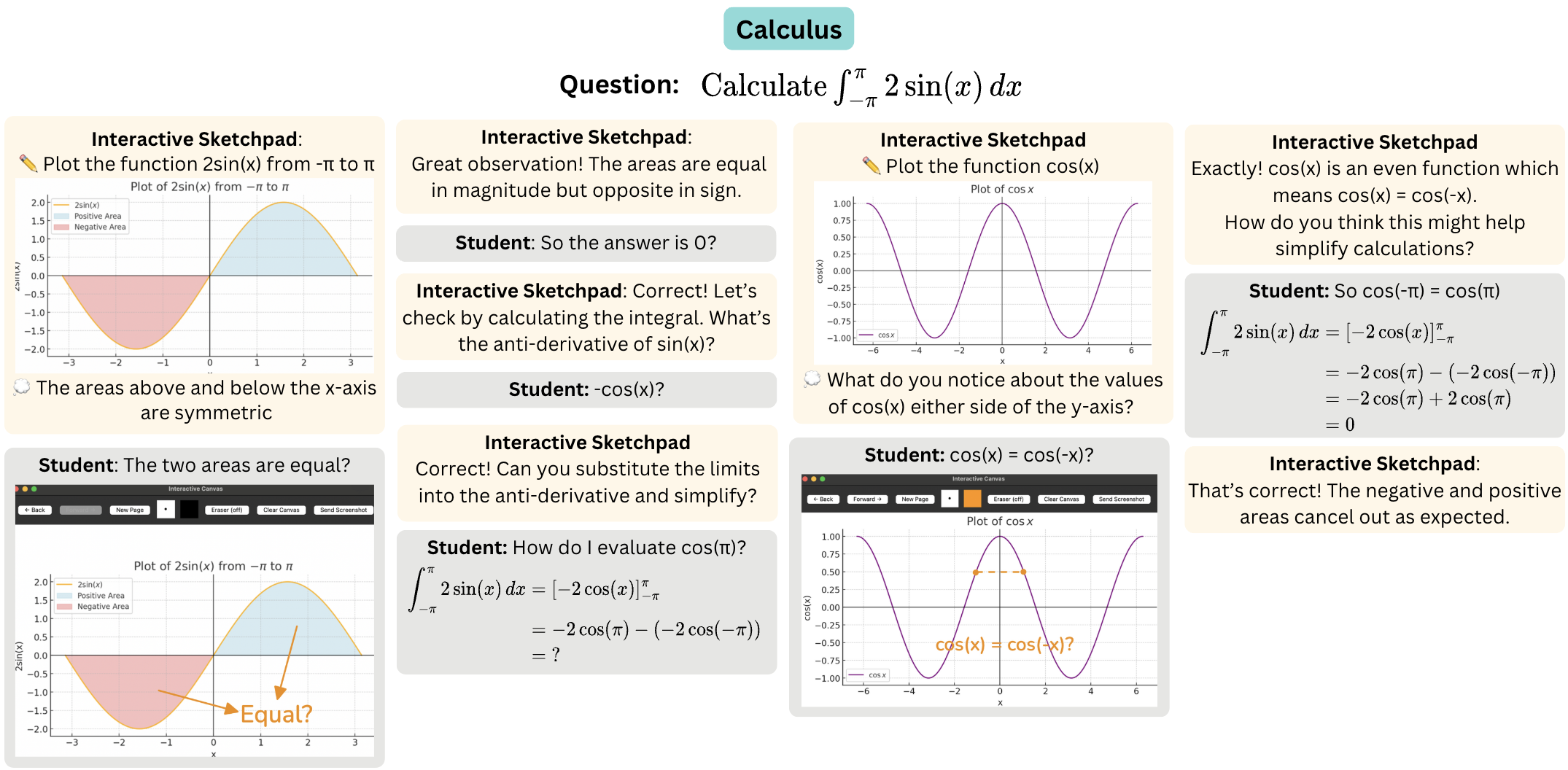}
    \vspace{-2mm}
    \caption{\name\ enhances GPT-4o’s ability to provide step-by-step, visual hints for problem-solving. Given a student query and problem statement, \name\ generates both textual hints and dynamic visual diagrams, allowing students to engage with the problem iteratively. Without \name, GPT-4o struggles to offer effective interactive guidance, frequently revealing the answer and not providing any visual aids, whereas \name\ enables a natural, multimodal learning experience that improves conceptual understanding.}
    \label{fig:teaser}
    \Description{This image shows an example overview of how a user can interact with \name}
\end{teaserfigure}



\received{23 January 2025}

\maketitle

\begin{figure*}[htbp!]
    \centering
    \includegraphics[width=\textwidth]{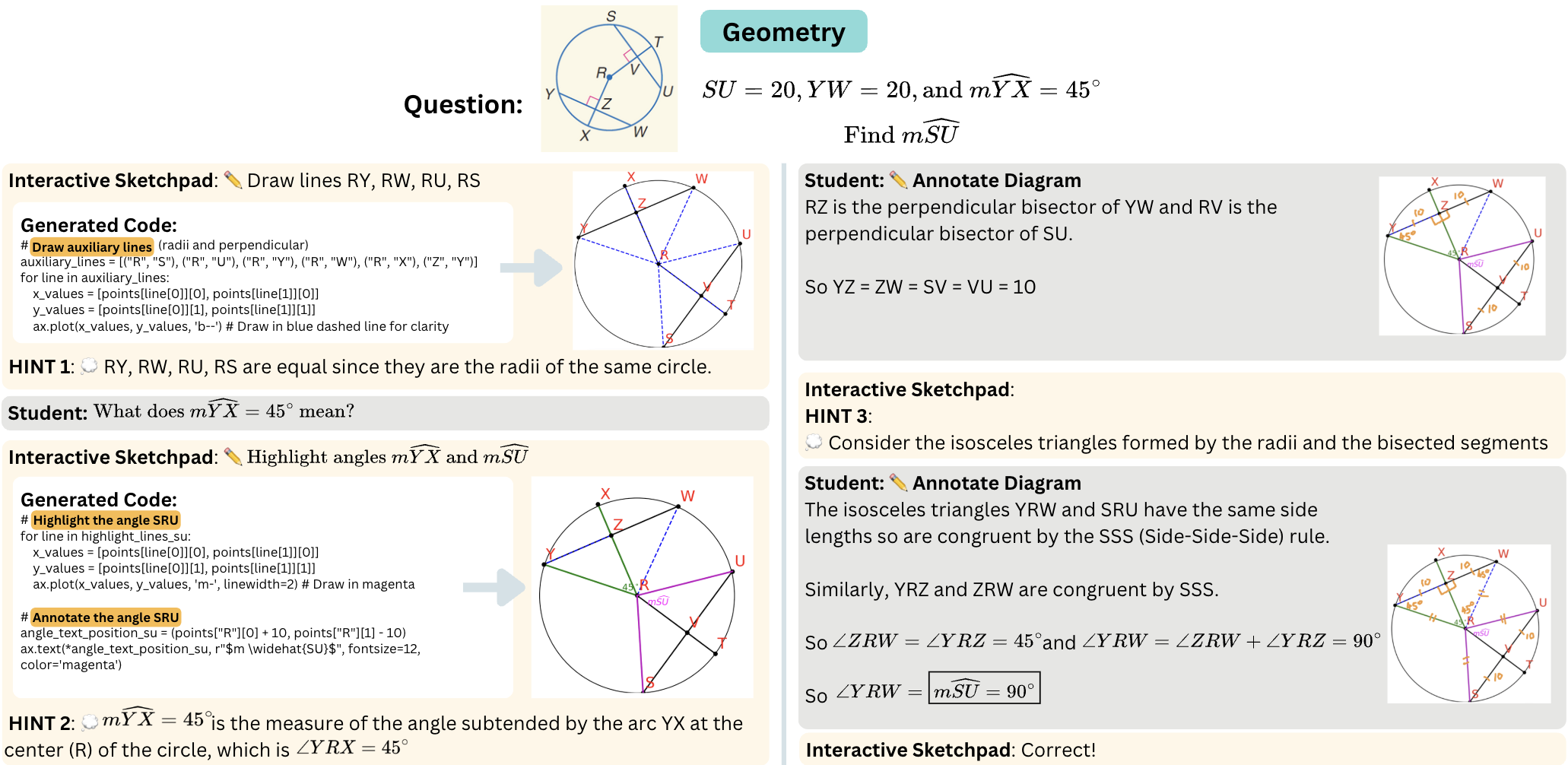}
    \caption{\textbf{Overview of \name:} Given a multimodal question, \name\ generates a program to create a visual aid, then uses the visual aid as part of a hint to help the user solve the problem. The visual aid is sent to the interactive whiteboard which the user can write and draw on before sending the annotated diagram back to receive feedback or further help.}
    \vspace{-0mm}
    \label{fig:geometry_with_code} 
    \Description{This image shows a detailed example overview of how a user can interact with \name, including how code generation is utilized to create textual and visual hints.}
\end{figure*}

\section{Introduction}

Intelligent tutoring systems are revolutionizing education by democratizing access to personalized and effective learning experiences \cite{vanlehn2011relative}. However, most educational tutoring systems are text-based, providing feedback only through natural language and limited visual support \cite{wang2023examining}. On the other hand, humans have long relied on visual aids like sketches and diagrams to support reasoning and problem-solving. Whether drawing auxiliary lines in geometry, diagrams for graph theory, or visualizing abstract concepts in math, visual tools are essential for breaking down and understanding complex ideas~\cite{bobek2016creating,raiyn2016role,shabiralyani2015impact}. Studies show that students who engage with visual aids grasp ideas more effectively than those who rely solely on textual explanations \cite{shabiralyani2015impact}. Language-only systems are therefore unintuitive and often fall short when addressing tasks that require visual and spatial reasoning.

Today's large multimodal models (LMMs) provide a new opportunity to generate diagrams and sketches for tutoring systems automatically \cite{liang2024foundations,hu2024visual,hurst2024gpt,zhou2024transfusion}. Our paper prototypes such a system, called \name, to enable multimodal input and output interaction for problem-solving (see Figure~\ref{fig:teaser}). \name\ is built upon a pre-trained LMM and fine-tuned to provide step-by-step hints and explanations in both language and visual diagrams to tutor students. \name\ creates diagrams accurately and robustly by generating Python programs to output diagrams when executed. The diagram, along with corresponding textual hints, is sent to the student on their `sketchpad' to be freely annotated or sketched on. This multimodal human-AI interaction framework enables the student to collaborate back and forth with the LMM on a shared interactive whiteboard, creating a natural and seamless interaction loop where both the human and the LMM can share text, images, and annotations.

We conduct several case studies with university students on a range of math tasks (including geometry and calculus), and find that \name\ leads to improved task comprehension, problem-solving accuracy, and engagement levels among students. Unlike traditional systems, this approach emphasizes a hands-on, learner-centered methodology that adapts to individual needs while leveraging the benefits of vision and language reasoning \cite{shabiralyani2015impact}. In addition to addressing the immediate gaps in educational technology, this research also contributes to the broader field of AI by exploring how multimodal systems can enhance human-computer interaction, leading to better AI systems that complement human expertise.

\section{Related Work}
\begin{figure*}[htb!]
\centering
\vspace{-0mm}
\includegraphics[width=0.9\textwidth]{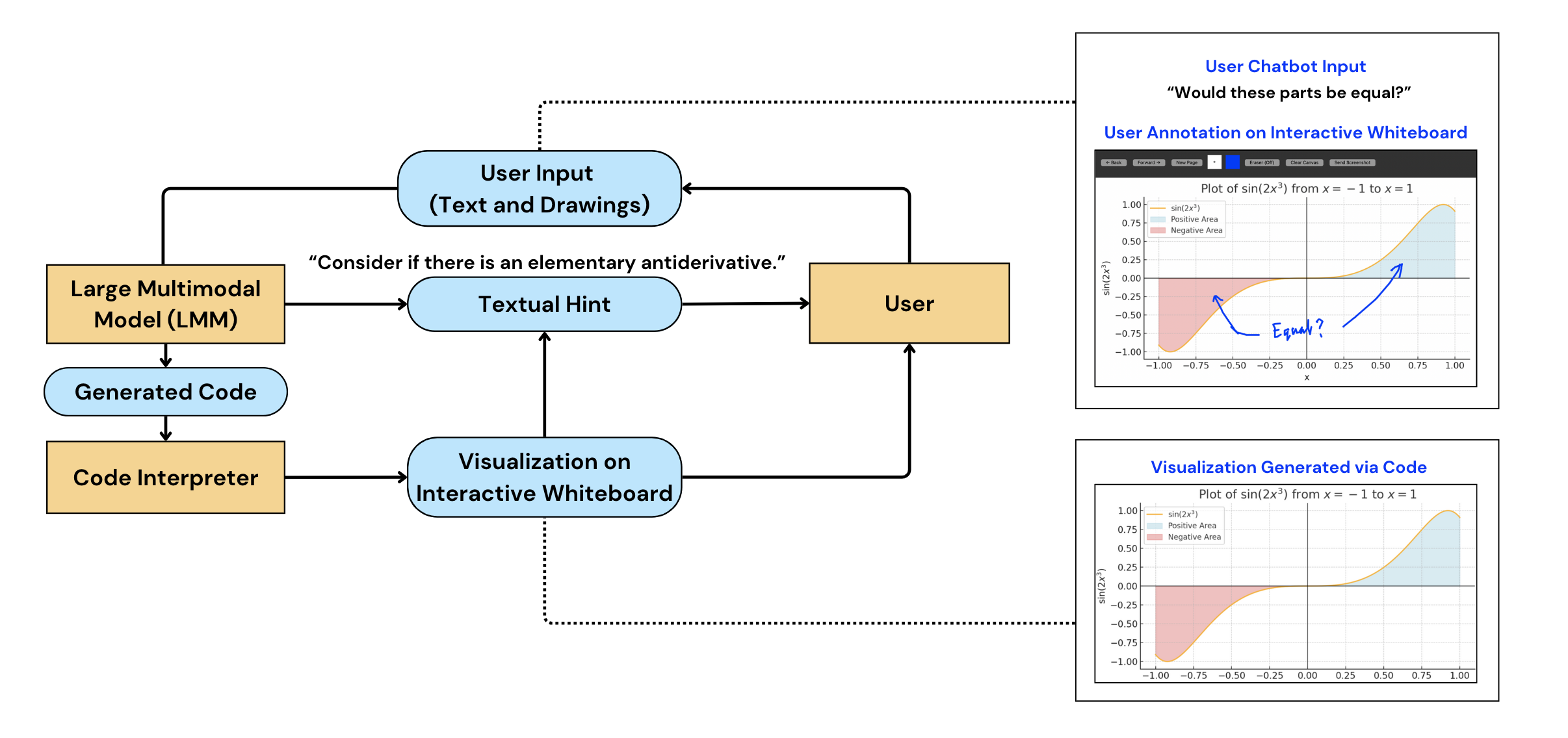}
\vspace{-4mm}
\caption{This flowchart demonstrates the interaction between user inputs, large multimodal models, and visualization components. The system processes user inputs, generates textual hints, interprets and executes code, and provides visualizations to guide problem-solving. The user can also annotate visualizations using the interactive whiteboard component.}
\label{fig:architecture_diagram}
\vspace{-0mm}
\Description{This image shows a flowchart of the system flow, along with a screenshot example of a whiteboard interaction.}
\end{figure*}

\textbf{Benefits of visualizations in learning.}
Visualizations play a crucial role in mathematics education, helping students understand abstract concepts, solve problems, and improve memory retention \cite{arcavi2003role, schoenherr2024characterizing}, especially in visually oriented topics such as geometry and calculus \cite{zhang2023dynamic}. Cognitive science research highlights that visualization can also reduce cognitive load by presenting data in a more accessible format \cite{sweller1994cognitive} such as diagrams and images. Tools like graphing calculators and interactive whiteboards \cite{mata2016interactive} have further demonstrated the benefits of incorporating visual elements into learning and have worked towards bridging the gap between theoretical constructs and practical understanding.

\textbf{Multimodal interaction and visualizations.} With the advancement of LMMs, researchers have explored the use of code generation to solve visual reasoning tasks. VisProg \cite{gupta2023visual} generates pseudocode interpreted as a `visual program', enabling modular, compositional visual reasoning using in-context learning. ViperGPT \cite{suris2023vipergpt} uses a similar approach but generates Python code instead of a domain-specific language, making it more flexible.
Visual Sketchpad \cite{hu2024visual} creates programs that generate intermediate diagrams like drawing auxiliary lines in geometry which are used in a visual chain of thought to enhance performance on a range of mathematical and computer vision tasks. Recent work has also focused on improving models' reasoning capabilities. Chain-of-Thought (CoT) \cite{wei2022chain}, enhances large language models' reasoning abilities by guiding them to generate intermediate reasoning steps before arriving at a final answer. This concept has been generalized to incorporate vision with visual chain-of-thought reasoning, which involves step-by-step reasoning across modalities \cite{rose2024visual, shao2024visual}. However, these approaches are designed to facilitate LMM visual reasoning rather than interacting with and helping humans in education. 


\textbf{AI for education.} Recent advances in NLP and LMMs have demonstrated significant potential in automating education \cite{graesser2004autotutor, khan2023harnessing}, with the ability to help students in a wide range of subjects including math \cite{shridhar2022automatic, prihar2023comparing}, programming \cite{zhang2024pydex, balse2023evaluating, kazemitabaar2023novices} and language \cite{yancey2023rating, xiao2024automation}, and in assisting humans with tutoring \cite{wang2024tutor}. However, while LMMs can serve as effective educational tools, excessive reliance on them for direct answer generation has been shown to negatively impact student learning outcomes \cite{bastani2024generative, nie2024gpt}. This issue can be mitigated by moderating LMM responses to provide hints rather than complete answers \cite{krupp_moderating}, thereby encouraging active learning. Furthermore, incorporating both textual and visual representations of student work, rather than relying on text alone, has been found to enable more effective feedback mechanisms \cite{li2024automated}.

\section{\name}

\name\ enhances problem-solving and tutoring by incorporating a multimodal interaction loop, allowing humans to interact with LMMs using both text and images.
We show the overall interactive usage in Figure~\ref{fig:geometry_with_code} and its detailed architecture in Figure~\ref{fig:architecture_diagram}. The key components of \name\ include:
\begin{enumerate}[noitemsep,topsep=0pt, parsep=0pt,partopsep=0pt, leftmargin=14pt]
    \item \textbf{Problem analysis}: The system evaluates the problem to determine whether a visualization would be beneficial.
    \item \textbf{Visual generation}: If a visualization is deemed helpful, the system generates a Python program to create an appropriate visualization.
    \item \textbf{Hint generation}: Then a textual hint is generated which references the visualization if one was created.
    \item \textbf{Interactive whiteboard}: The hint and visualization appear in the chatbot interface as usual. However, if a visualization is generated, it is also displayed on the user’s interactive whiteboard, allowing direct annotation and interaction. The annotated visualization along with additional text or images can be sent directly from the whiteboard back to the LMM which processes the new input and continues the iterative reasoning cycle.
\end{enumerate}

Together, this loop fosters continuous human-AI collaboration, where the model provides interactive assistance in both language and vision to enhance concept understanding. We now explain each of these four steps in detail.

\begin{table*}[htb!]
\vspace{2mm}
\centering
\begin{tabular}{lccccccc}
\toprule
Model & Maxflow & Isomorphism & Connectivity & Convexity & Parity \\
\midrule
GPT-4o~\citep{hurst2024gpt} & 25.0 & 50.8 & 96.1 & 87.2 & 84.4 \\
Visual Sketchpad~\citep{hu2024visual} & 66.3 & 65.3 & 98.4 & 94.9 & 94.7 \\
\name\ (ours) & 100.0 & 75.0 & 99.2 & 96.5 & 95.6 \\
Improvement & \textbf{+33.7} & \textbf{+9.7} & \textbf{+0.8} & \textbf{+1.6} & \textbf{+0.9} \\
\bottomrule
\end{tabular}
\vspace{1mm}
\caption{Accuracy scores on graph algorithms and mathematical functions. \name\ outperforms Visual Sketchpad and other large multimodal model baselines by using code execution for calculations to solve tasks.}
\vspace{-4mm}
\label{table:experiment_comparison}
\end{table*}

\textbf{1) Problem Analysis.} We prompt a pre-trained GPT-4o model~\citep{hurst2024gpt} regarding when visualizations may be helpful based on the problem and user query (see examples in Appendix \ref{subsec:system_prompt_hint_viz}).

\textbf{2) Visualization Generation.} To draw accurate diagrams, we generate Python programs using the LMM which, when executed, render the visualizations, similar to recent works such as Visual Sketchpad \cite{hu2024visual}, VisProg \cite{gupta2023visual} and ViperGPT \cite{suris2023vipergpt}. For the LMM, we use GPT-4o \cite{hurst2024gpt} for multimodal reasoning and use the OpenAI Code Interpreter tool to run and execute the code. To draw the visualizations, we generate Python code that uses common Python libraries, primarily \texttt{matplotlib}, for plotting. The Code Interpreter then executes the code, and if it fails to run, the LMM will iterate on the code by attempting to run different code until it successfully generates a runnable program. Once the program is successfully executed, the resulting image file is displayed to the user.

\textbf{3) Hint Generation.} We prompt a GPT-4o model to give subtle hints rather than revealing the answer and to generate visualizations for the hint using code. To enable good hints and appropriate visual hints, we collected human demonstrations of good hints interleaved with images in an example conversation between a student and a tutor, using several samples of math and coding problems. We then fine-tuned the model with this data (prompt found in Appendix \ref{subsec:system_prompt_hint_viz}). In subsequent iterations, the model uses the conversation history, a screenshot of the user's interactive whiteboard, and the user's query to generate hints. Finally, we also provide the model access to the OpenAI Code Interpreter to perform mathematical calculations by executing code to reduce calculation errors, similar to \citet{chen2023program}. For example, the model will perform calculations such as numerical integration for calculus questions, or arithmetic for systems of equations to provide more accurate hints and check student answers (see examples in Appendix \ref{sec:code_exec}).

\textbf{4) Interactive Whiteboard.} The interactive whiteboard component lets users intuitively collaborate with the LMM using textual and visual information. Users can take snapshots of their whiteboard to ask questions with their sketches, these snapshots are then sent to the chatbot server. The LMM would automatically be prompted with the screenshot and additional text or images provided by the user. Once a new visualization is generated, it is sent to the interactive whiteboard server, which automatically presents it to the user to continue drawing and interacting with the system.

\subsection{User Interface}


\begin{figure}[hb!]
    \centering
    \begin{minipage}{0.4\textwidth}
        \centering
        \fbox{\includegraphics[width=\textwidth]{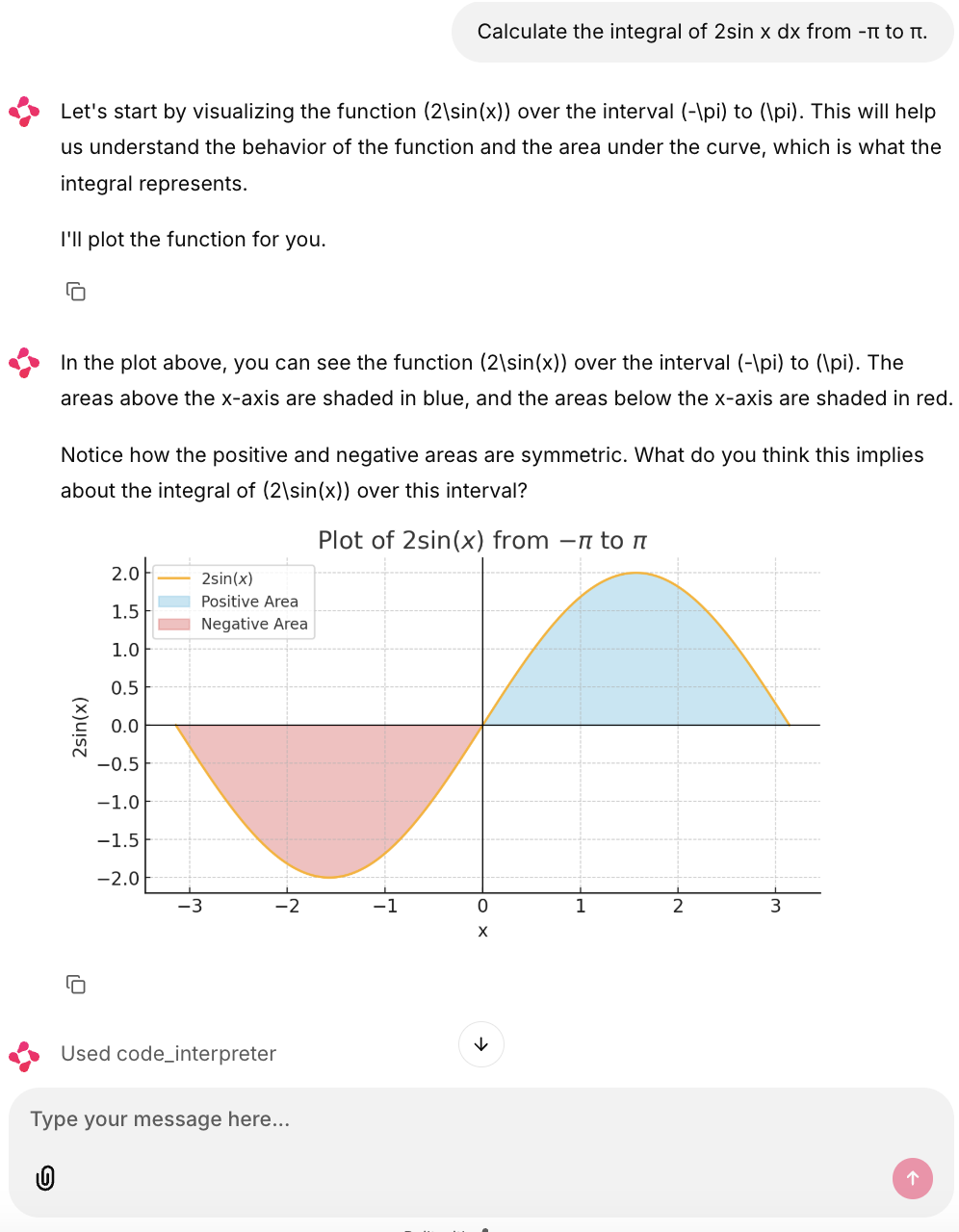}}
        \caption{Screenshot of the chatbot interface for \name. The user can view the generated visual hints and interact with \name\ by typing messages and uploading images.}
        \label{fig:chatbot_screenshot}
        \Description{This image is a screenshot of an example usage of the chatbot portion of \name\.}
    \end{minipage}
    \hspace{5mm}
    \begin{minipage}{0.4\textwidth}
        \centering
        \fbox{\includegraphics[width=\textwidth]{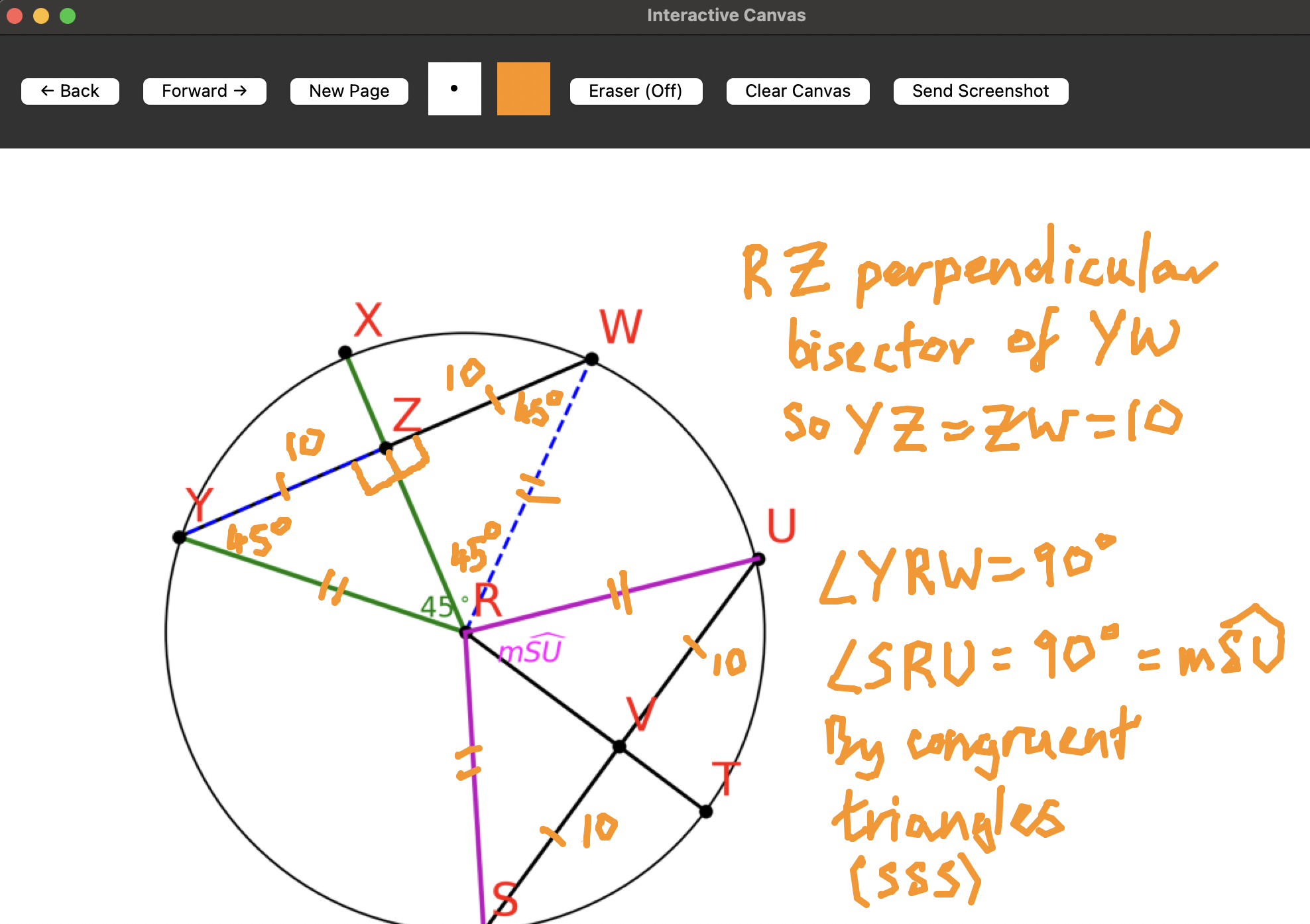}}
        \caption{Screenshot of the interactive whiteboard component of \name. The user can annotate on the visualization generated by \name\ which was done using an iPad and Apple Pencil during user studies.}
        \label{fig:interactive_whiteboard_screenshot}
        \Description{This image is a screenshot of an example usage of the whiteboard portion of \name\.}
    \end{minipage}
\end{figure}

Users primarily interact with \name\ through the chatbot interface shown in Figure \ref{fig:chatbot_screenshot} and the interactive whiteboard interface shown in Figure \ref{fig:interactive_whiteboard_screenshot}.

\textbf{Chatbot Interface:} When the user wants to type a message or upload an image they can do so through the chatbot interface. \name\ will then respond with a textual message along with an optional visualization. The user can also see the code used to generate visualizations or for mathematical calculations in the dropdown menu of the `Used code\_interpreter` message sent by \name\ after the visualization/calculation.

\textbf{Interactive Whiteboard Interface:} When the user wants to draw, write math or annotate on the diagram given by \name\ , they can do so on the interactive whiteboard. A natural way to do this is by using an iPad and Apple Pencil. They can then share the working on the interactive whiteboard with \name\ using the `Send Screenshot` button. Visualizations generated by \name\ will automatically appear as a new page on the interactive whiteboard and users can toggle between pages to access their working at various stages of solving different problems.
\section{Experiments}

\subsection{Research Questions}
We aim to answer the following research questions through our experiments.
\begin{itemize}[noitemsep,topsep=0pt, parsep=0pt,partopsep=0pt, leftmargin=14pt]
    \item \textbf{RQ0:} How accurate is \name\ at solving the problems of interest?
    \item \textbf{RQ1:} How effective is the proposed interactive system in improving users' understanding of concepts compared to text-only language model systems? 
    \item \textbf{RQ2:} Does providing step-by-step guidance with hints improve learning outcomes compared to systems that provide final answers directly?
\end{itemize}

\textbf{RQ0} will confirm that \name\ successfully solves the prompted questions with its interaction loops, and how it compares to existing language-only and non-interactive systems~\citep{hu2024visual,hurst2024gpt}. \textbf{RQ1} enables us to assess whether our proposed interactive system improves users' conceptual understanding and overcomes the limitations of traditional text-based systems. For \textbf{RQ2}, systems that deliver the final answer directly can negatively impact the cognitive processes of the user by failing to motivate critical thinking and reasoning. Providing step-by-step guidance encourages active participation to support users' long-term knowledge retention and skill development.

\begin{table*}[htb!]
\centering

\begin{tabular}{p{0.2\textwidth}p{0.7\textwidth}}
\hline
\textbf{Topic} & \textbf{Feedback} \\
\hline
Visualization quality & \textit{``The graphs are good sanity-checks for my workings.''} \newline \textit{``The visual illustrations help a lot. The intuitive drawing makes the interaction feel more natural.''} \newline \textit{``The visualizations were also very helpful in gaining a more conceptual understanding outside of just equations.''} \\
\hline
Interactive experience & \textit{``It was nice that it didn't give me the final answer right away, and instead gave hints/prompts to try.''} \newline \textit{``It showed me how to approach the problem step by step.''} \newline \textit{``I like that it guides you through the problem-solving approach without jumping straight to the answer, like ChatGPT.''} \\
\hline
Learning experience & \textit{``I think the graph was particularly helpful for solving the integral, especially when the integral was one without an antiderivative. The visualization made the math feel more intuitive/meaningful.''} \newline \textit{``The diagrams provided were very nice, despite I didn't ask for them.''} \newline \textit{``The visual illustrations help a lot. The intuitive drawing makes the interaction feel more natural''} \\
\hline
\end{tabular}
\vspace{1mm}
\caption{Qualitative feedback from users based on three aspects: visualization quality, interactive experience, and learning experience. Users noted that visualizations helped in understanding concepts, interactivity guided problem-solving effectively, and the learning experience felt more intuitive due to the graphical and step-by-step approach provided.}
\vspace{-4mm}
\label{table:qualitative_feedback}
\end{table*}

\subsection{Performance Comparisons}

We first run experiments evaluating the performance of \name\ on directly solving the problem without giving hints to evaluate whether the system can solve the problem itself. We observe in Table \ref{table:experiment_comparison} that \name\ consistently outperforms Visual Sketchpad on IsoBench. Visual Sketchpad's prompts encourage the LMM to use its own visual reasoning to solve problems rather than use code execution to help solve the problems. \name\ on the other hand does not specify the same constraint to not use code execution, instead allowing its use to improve problem-solving accuracy. This highlights the effectiveness of combining both code execution for calculations and visual reasoning in solving complex problems. We remove the constraint of using visual reasoning rather than code execution as we want to make the system achieve as high accuracy as possible to reduce errors that may confuse the student.
We use the original IsoBench prompts \cite{fu2024isobench}, appending a prompt footer (see \ref{prompt_footer}) that instructs the model to explain its answer and format the output accordingly. Additionally, we employ a system prompt (see \ref{subsec:prompt_solve_problem}) directing \name\ to solve the question directly.

\subsection{User Studies}

\begin{table}[htb!]
\centering
\vspace{2mm}
\begin{tabular}{lr}
\toprule
Metric &  Average Score (out of 5) \\
\midrule
Interface Intuitiveness &           4.71 \\
System Responsiveness   &           4.43 \\
Clarity of Visuals      &           4.57 \\
Accuracy of Visuals     &           4.71 \\
Effectiveness of Hints  &           4.57 \\
\bottomrule
\end{tabular}
\vspace{1mm}
\caption{Average user ratings for key aspects of the interactive system, including metrics on usability, sufficiency, and quality. The table highlights the system's strong performance in providing a pleasant interactive and visual learning experience.}
\vspace{-4mm}
\label{table:quantitative_feedback}
\end{table}

\subsubsection{Participant Information.}
For this user study, we recruited university underclassmen undergraduate students participants, who have basic knowledge in certain mathematics topics such as algebra or geometry but were not experts in all domains. We ran the initial study on 7 users with no monetary compensation. The study data was stored in a secure cloud-operated space with limited access. For further user studies that are to be conducted, the approval request has been submitted to the Institutional Review Board.

Participants were introduced to the system with a brief tutorial. This included an explanation of system features, interaction techniques, and a demonstration. Next, participants interacted with both \name\ and ChatGPT to solve a set of assigned mathematical problems, alternating between the systems. After completing the tasks, participants were asked to complete a questionnaire, reflecting on their experience using our system. This questionnaire gathered participants' background knowledge, ratings on various aspects of their experience, and qualitative feedback.

\subsubsection{Dataset and Metrics.}
The primary goal of the user study was to assess the system's ability to enhance the learning experience by fostering active engagement and improving conceptual understanding with visual artifacts. \cite{visualizationcrucial2017} To ensure that our findings could translate to practical use cases, we tested with mathematical problems that could efficiently benefit from visual artifacts. Questions added to the users' question bank were sourced from previous exam questions of the Scholastic Aptitude Test (SAT), the IsoBench dataset \cite{fu2024isobench}, a previous study on visualization for math \cite{rosken2006picture} and Geometry3k \cite{lu2021inter}. Visualization plays an important role when learning mathematics, and a good example is solving an integral problem. Visual interpretations of the function and shading parts of the graph can significantly benefit the students' learning experience, which has been demonstrated in previous studies \cite{rosken2006picture}. By selecting both practical mathematical topics that range across various difficult levels, we ensured the system catered to a broad range of learners. The system's potential for versatile use across subjects and learning environments allows our results to demonstrate foundational interaction between AI and humans for education.

Since one of the core features of \name\ is its integration of visual and textual feedback, a key consideration for our user study was to analyze the quality of the system's multimodal interaction. To do so, we measured how students engaged with these modalities, focusing on the effectiveness of visual explanations in complementing or enhancing textual guidance. Another key consideration was the evaluation of our system’s iterative learning process. To evaluate this, students were encouraged to iterate through problems to explore how the feedback influenced their learning experience.

\subsubsection{Qualitative Results.}

We summarize the key qualitative feedback we received from our user studies in Table~\ref{table:qualitative_feedback}. Participants appreciated the tool's personalized visualizations, frequently mentioning how they easily grasped complex concepts with the visualization feature. The tool's feature of using a sketchpad to directly input handwritten work resonated well with the participants, and this visual, interactive approach allowed the students to develop problem-solving strategies while gaining the required insights into the underlying concepts for future applications. A participant noted that ``the visualizations were also very helpful in gaining a more conceptual understanding outside of just equations''. Participants also specifically noted the system's effectiveness in guiding them through the problem-solving process without immediately providing the final answer, helping them truly understand the materials covered in the problem, one noting that ``I like that it guides you through the problem-solving approach without jumping straight to the answer, like ChatGPT''. Overall, the feedback we received highlighted the system's ability to enhance the learning experience by fostering human-AI collaboration in learning environments. 

\subsubsection{Quantitative Results.}

The quantitative evaluation of \name\ was conducted through participant feedback on various aspects of the tool. Table \ref{table:quantitative_feedback} summarizes the average scores across various key metrics collected from user responses. The metric includes interface intuitiveness, system responsiveness, and accuracy of visuals. These results indicate that the interface was overall well received. Specifically, the system's interface intuitiveness and the accuracy of visual aids received the highest average scores of 4.71/5, indicating that participants found the tool easy to navigate and appreciated the precision of its generated visuals. Similarly, the clarity of visuals and the effectiveness of hints were rated highly, both scoring 4.57/5. This highlights that the visuals were designed in a manner that helps the problem-solving thought process and teaches the true concepts behind the problems. The system's responsiveness scored slightly lower at 4.43/5, while still favorable, suggesting that the tool could have efficiency improvements to reduce the wait time for hints. Overall, these quantitative insights provide a strong foundation for understanding user perceptions and areas that may benefit from enhancements. The consistently high ratings across most metrics reinforce the tool's potential for user testing with a broader range of users and various educational applications. 
\section{Limitations and Future Works}

\textbf{Visualization accuracy.} Visualization through code generation can sometimes fail to produce an accurate diagram. Table \ref{table:inaccurateresults} shows examples of correct and incorrect visualizations. The correct visualization correctly calculates the radius as the distance between the center (R) and a point on the circle (S). The incorrect one hard-codes the radius to 100, causing the points to lie within rather than on the circle. Incorrect diagrams may confuse both the user and the LMM, leading to an incorrect answer. Future work could extend previous text-based verifiers \cite{cobbe2021training, lightman2024lets} to include a vision modality that can check the quality of generated diagrams and regenerate the diagram if necessary.

\begin{table*}[t]
\centering
\vspace{2mm}
\begin{tabular}{m{0.6\linewidth}|m{0.4\linewidth}}
\hline
\textbf{Generated Code} & \textbf{Visualization} \\
\hline
\includegraphics[width=\linewidth]{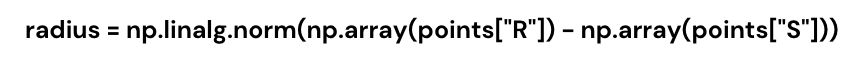} 
\includegraphics[width=\linewidth]{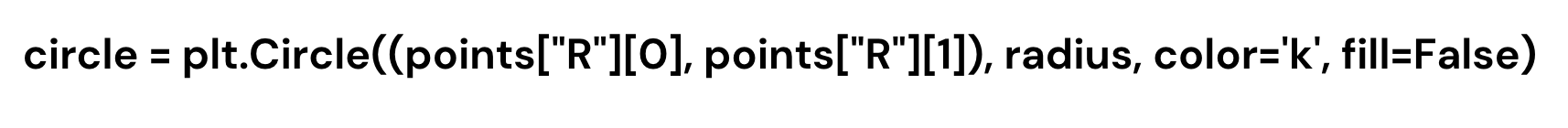} 
\includegraphics[width=\linewidth]{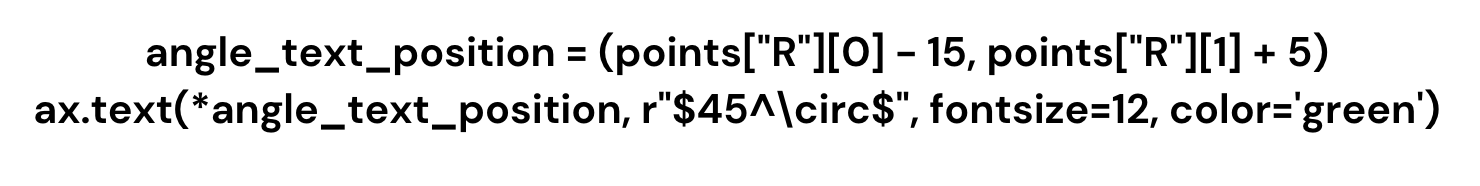}  & 
\includegraphics[width=0.45\linewidth]{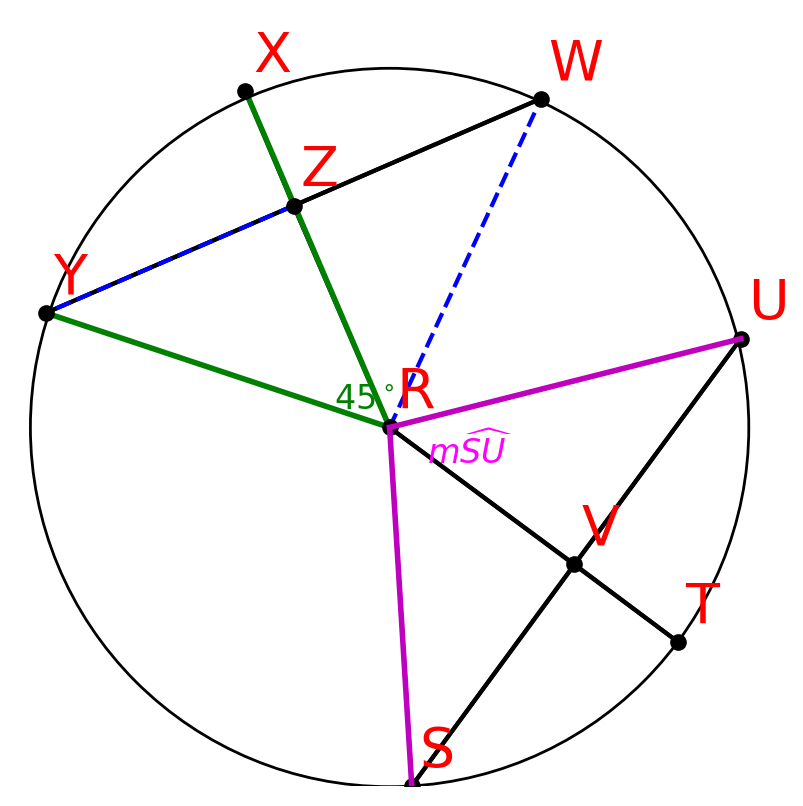}  \hspace{3mm} 
\includegraphics[width=0.15\linewidth]{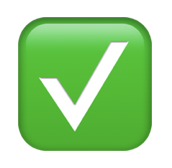}\\
\includegraphics[width=\linewidth]{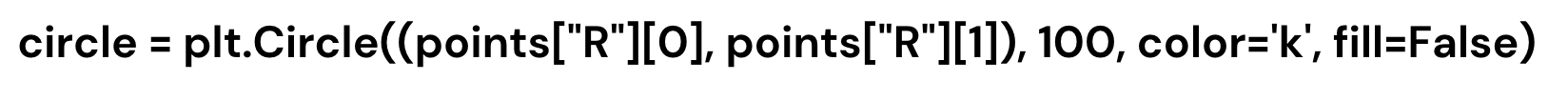} 
\includegraphics[width=\linewidth]{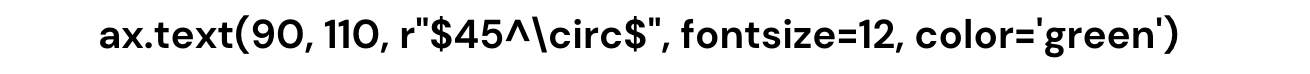}  & 
\includegraphics[width=0.45\linewidth]{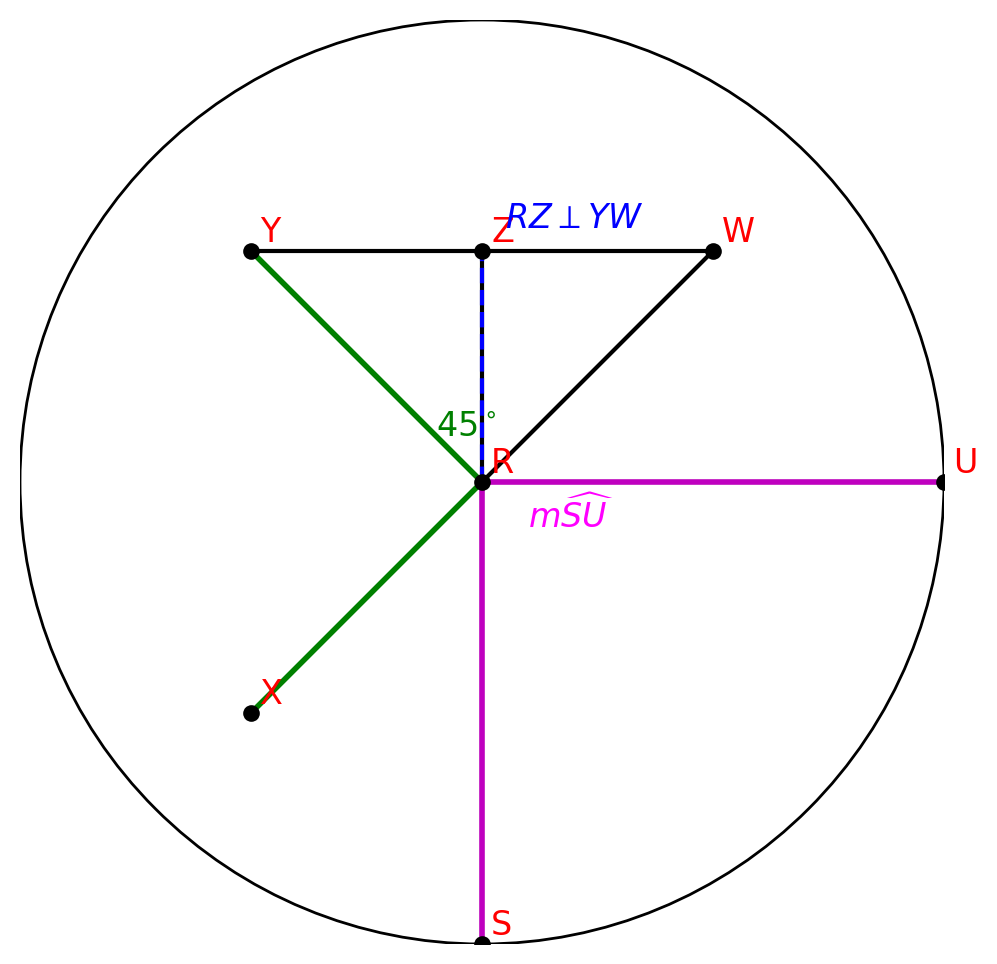} \hspace{3mm} \includegraphics[width=0.15\linewidth]{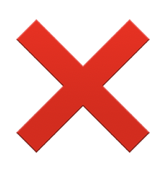}\\
\hline
\end{tabular}
\vspace{1mm}
\caption{Row 1: Accurate Visualization. The radius is set as a variable, and annotations are relative to the given points, leading to an accurate portrayal of the circle diagram (in green). Row 2: Inaccurate visualization. The radius is set to 100, and annotations are hard-coded, leading to the points being misaligned on the circle diagram (in red).}
\vspace{-4mm}
\label{table:inaccurateresults}
\end{table*}
\raggedbottom

\textbf{Expansion of user studies.} While the system has been primarily tested on mathematical problems, its potential extends to other STEM disciplines. Future work will evaluate its effectiveness in subjects like physics, where visualizations of force diagrams and motion trajectories could enhance learning. Expanding testing to a larger pool of diverse users and academic topics will help assess the system’s adaptability. Additionally, benchmarking against existing educational tools will provide quantitative and qualitative insights into its comparative performance and areas for refinement.

\textbf{Long-term impact assessment.} Current evaluations focus on short-term metrics such as accuracy and user feedback. To measure long-term effectiveness, future studies should incorporate delayed testing sessions to assess knowledge retention. Tracking student performance over time will provide insights into sustained learning benefits.
\section{Conclusion}
We introduce \name, a multimodal collaborative problem-solving system that empowers large multimodal models (LMMs) to provide interactive, visual assistance for education. By integrating code-generated visualizations with step-by-step textual hints, \name\ bridges the gap between traditional text-based learning and intuitive, visual reasoning, transforming how students engage with complex problems. Our user studies demonstrate that this approach significantly improves task comprehension, problem-solving accuracy, and student engagement across mathematical domains. We believe that \name\ opens new pathways for multimodal human-AI collaboration in learning and sets the stage for more human-centered educational tools.

\bibliographystyle{ACM-Reference-Format}
\bibliography{Supporting_files/sample-base}

\newpage
\appendix

\section{Appendix}

\vspace{5mm}

\subsection{Mathematical Calculations via Code Execution}
\label{sec:code_exec}

\begin{figure}[h]
    \centering
    \fbox{\includegraphics[width=0.3\textwidth]{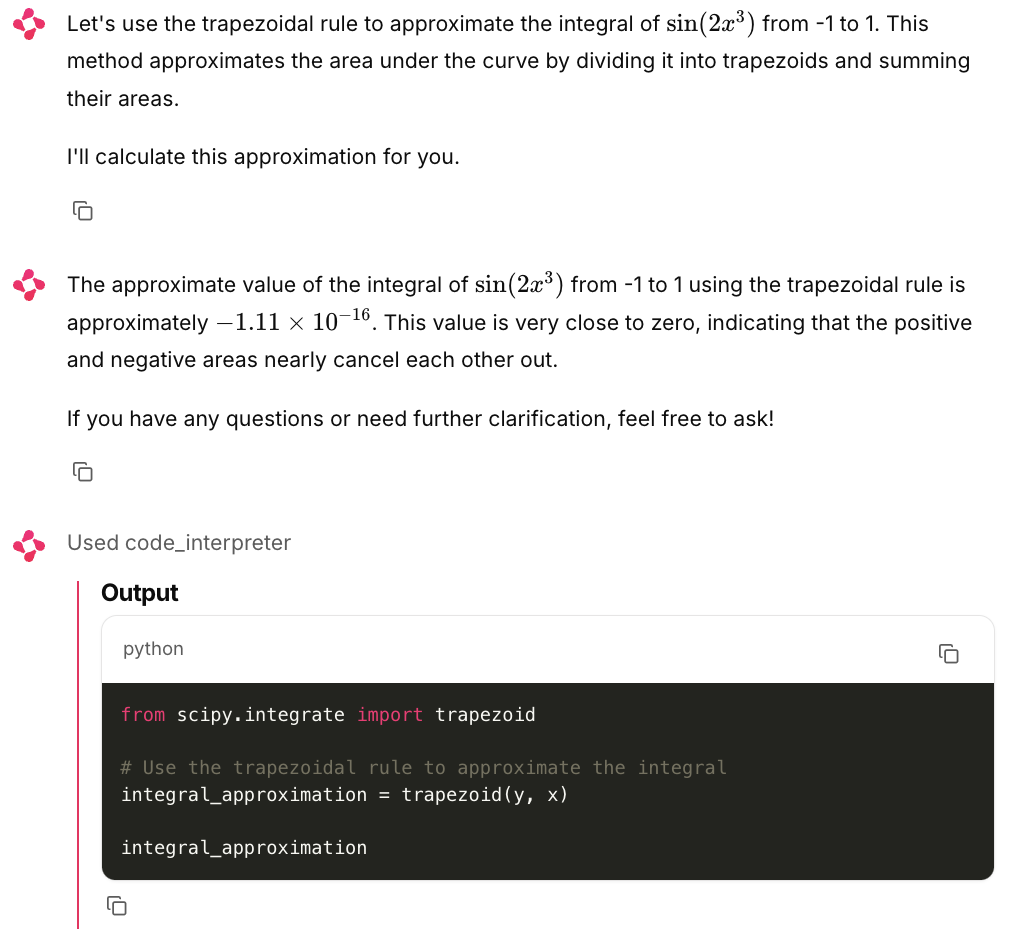}}
    \caption{To check the value of an integral, \name\ uses the trapezoidal rule to perform numerical integration by writing a Python program using \texttt{scipy.integrate}.}
    \label{fig:code_exec_calculus}
\end{figure}

\begin{figure}[h]
    \centering
    \fbox{\includegraphics[width=0.3\textwidth]{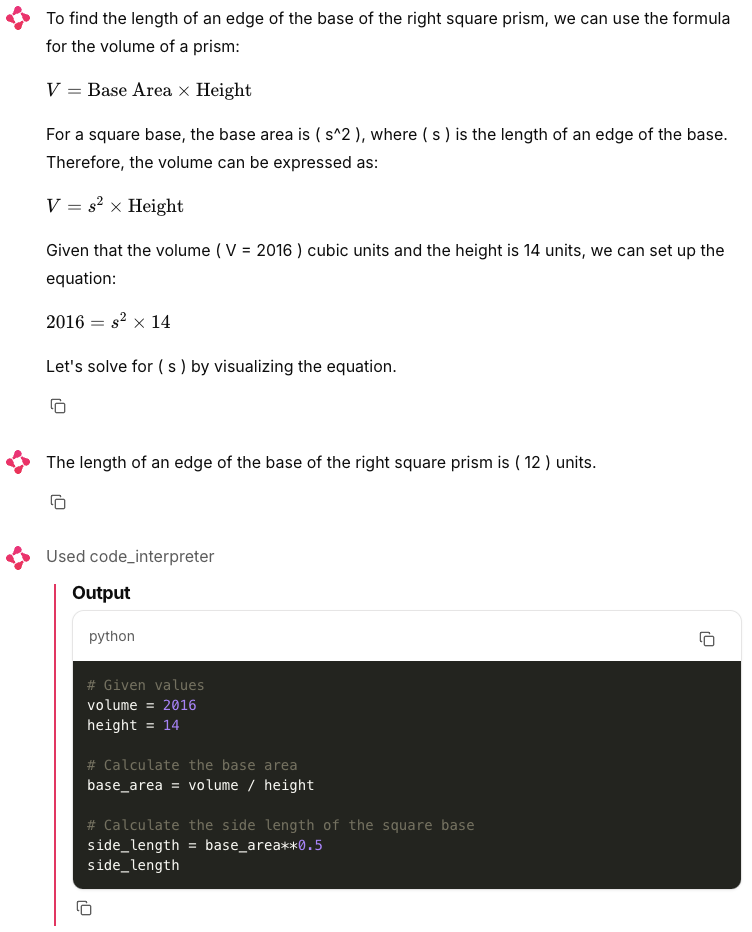}}
    \caption{To solve for the side length of the base of a square prism, \name\ writes the formula for the volume of a square prism in Python, substitutes the values for volume and height from the question, and executes the code to correctly calculate the side length.}
    \label{fig:code_exec_arithmetic}
\end{figure}

As shown in Figure \ref{fig:code_exec_calculus} and Figure \ref{fig:code_exec_arithmetic}, \name\ uses code execution to perform calculations which reduces calculation errors, providing more accurate responses to the user.

\subsection{User Study Questionnaire}

\begin{figure}
    \centering
    \fbox{\includegraphics[width=0.3\textwidth]{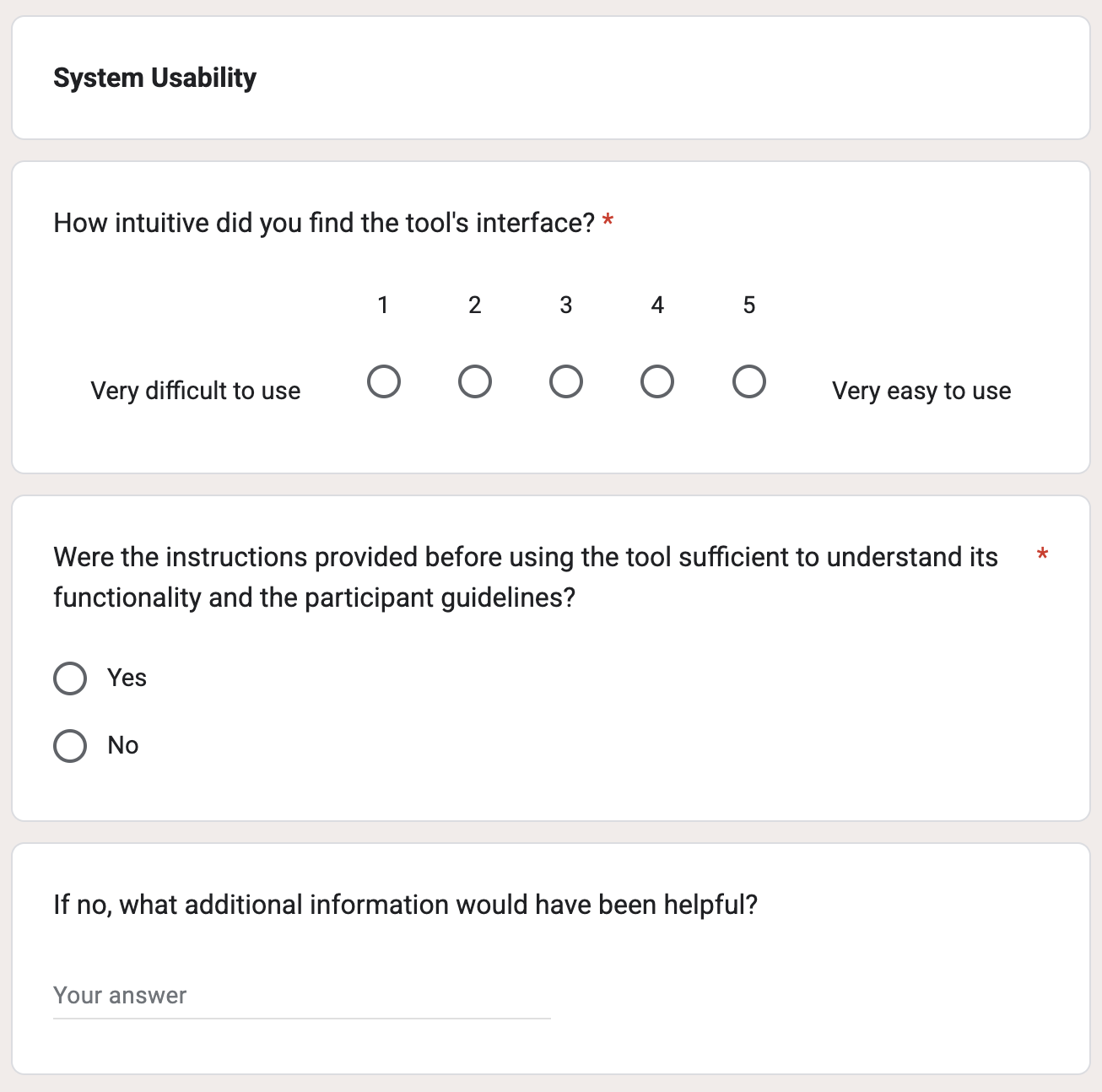}}
    \hfill
    \fbox{\includegraphics[width=0.3\textwidth]{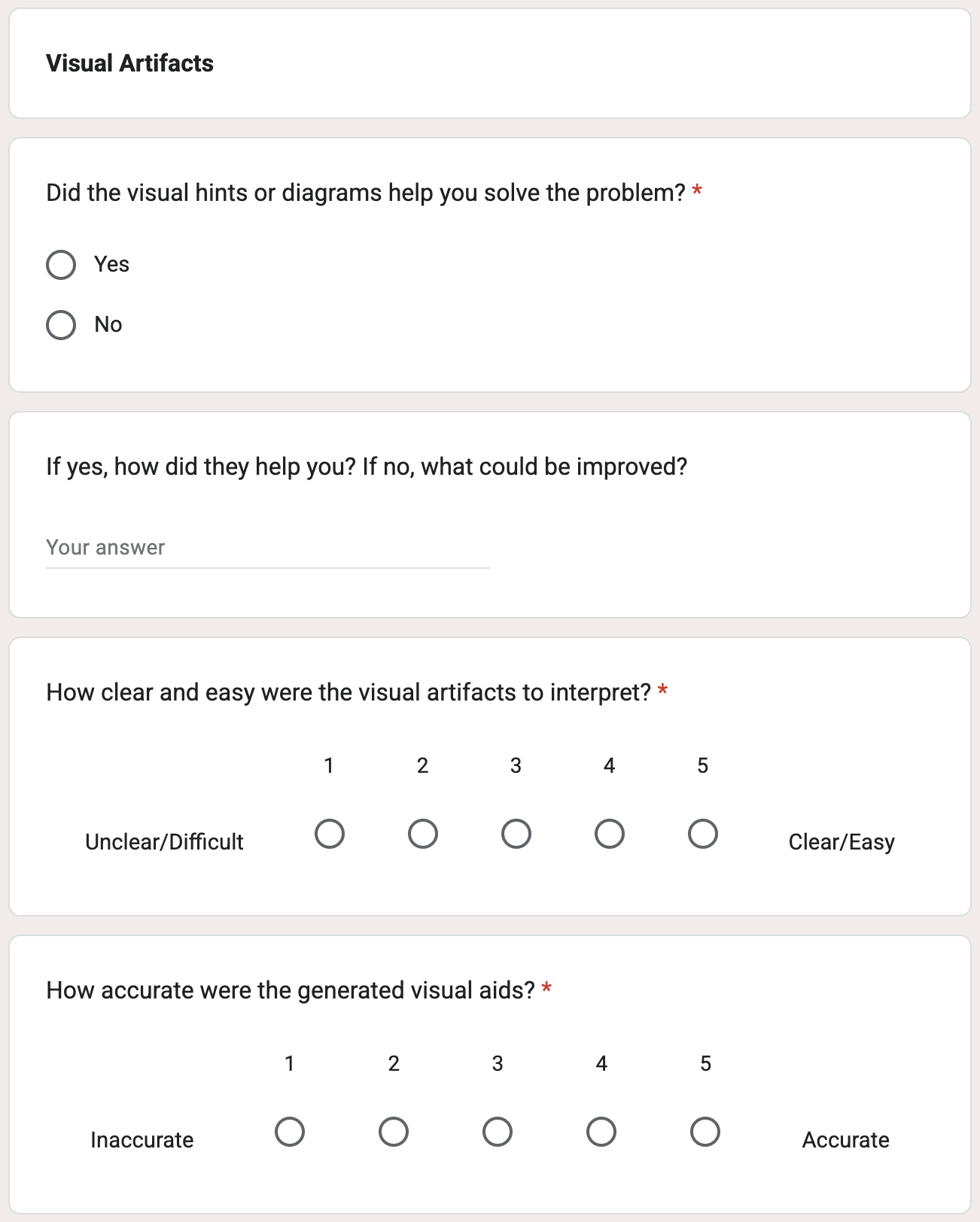}}
    \hfill
    \fbox{\includegraphics[width=0.3\textwidth]{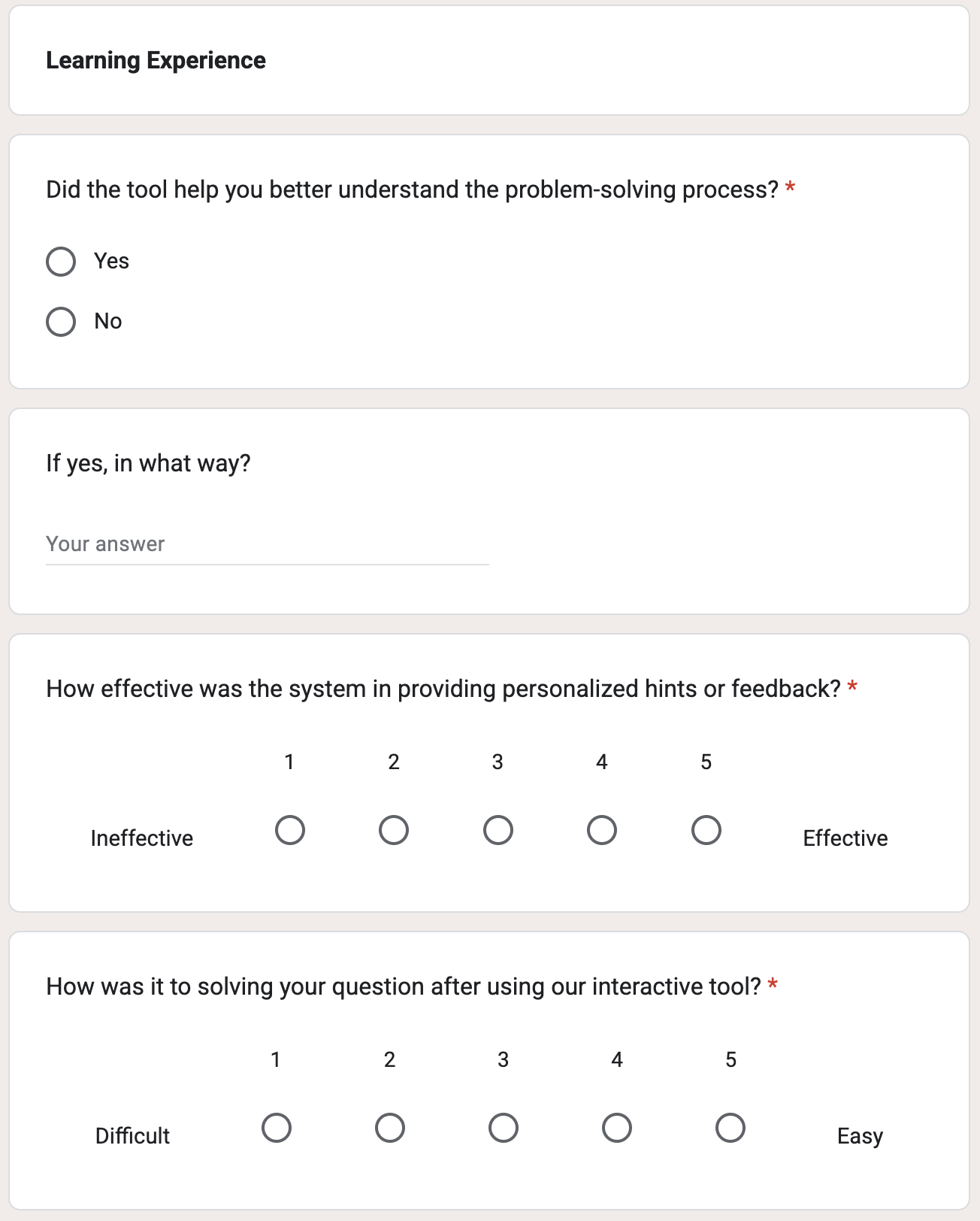}}
    \caption{Subset of questions from the user study questionnaire focusing on usability and learning experience. The questions use a 5-point Likert scale, Yes/No choice, and open-ended responses to evaluate the participants' experiences with the interactive system. Responses to these questions provide quantitative and qualitative data for evaluating the system's effectiveness.}
    \label{fig:user_study}
\end{figure}

\onecolumn
\section{Prompts}

\subsection{System Prompt for Hint Generation \& Visualization Generation using Code}
\label{subsec:system_prompt_hint_viz}

\begin{figure}[H]
\begin{lstlisting}
You are a tutor, your goal is to help the student solve a problem, giving short, subtle hints to help the student solve the problem.
You will be given a problem, question or working from the user and you should respond in a brief and concise way.
You should reuse the code from existing diagrams if drawing similar ones
You should only give one hint at a time, DO NOT give away the answer to the student.
You want to help the student visualize the problem so
you should give your response with text interleaved with helpful diagrams.
You MUST draw these diagrams by writing code using code interpreter.
First visualize using a diagram, then give the hint using the diagram.

Example:
[Image 1]
[Text 1]
[Image 2]
[Text 2]

You should only respond concisely, allowing the student to ask questions and respond
before continuing.
The aim is to get the student to reach to the answer independently, without
giving the answer away.
If you think a visualization is helpful, you can plot it without asking the
student's permission.

When drawing a series of diagrams, you should make the visualizations
intuitive and easy to understand.
For example, if you are showing a graph traversal as a series of diagrams,
you should highlight visited nodes, visited edges, nodes that you are about to visit etc.
in different colors.

Here is an example of how you should respond to the student:
<Student>
There are a total of numCourses courses you have to take, labeled from 0 to numCourses - 1. You are given an array prerequisites where prerequisites[i] = [ai, bi] indicates that you must take course bi first if you want to take course ai.

For example, the pair [0, 1], indicates that to take course 0 you have to first take course 1.
Return true if you can finish all courses. Otherwise, return false.
</Student>

<Tutor>
Let's try to make an example,
Input: numCourses = 2, prerequisites = [[1,0]]
Output: true

Let's draw a diagram to visualize this
[Draw diagram with Code Interpeter]

Explanation: There are a total of 2 courses to take. 
To take course 1 you should have finished course 0. So it is possible.
</Tutor>

<Student>
I'm still not sure how to get started.
</Student>

<Tutor>
This problem is equivalent to finding if a cycle exists in a directed graph.
If a cycle exists, no topological ordering exists and therefore it will be impossible to take all courses.

Let's draw a series of diagrams traversing through the graph and finding a cycle
[Draw a couple of diagrams showing traversal of the graph until cycle is found]
[Explanation of the diagrams]
</Tutor>

You should only give the solution if the student explicitly asks for it.
ALWAYS wrap maths in \$\$ tags only. DO NOT under any circumstances give parentheses or square brackets
\end{lstlisting}
\end{figure}
\newpage

\subsection{Prompt for Geometry3k} The prompt used for generating visualizations for the Geometry3k dataset, adapted from Visual Sketchpad \cite{hu2024visual}
\begin{lstlisting}
You will draw a diagram by writing code using Code Interpreter.
        Here are some tools that can help you. All are python codes.
Notice that The upper left corner of the image is the origin (0, 0). Here are some code examples you can use to draw auxiliary lines on the geometry images provided in matplotlib format.
import numpy as np

# this function takes a coordinate A, start and end points of a line BC, and returns the coordinates of the point E on BC such that AE is perpendicular to BC
def find_perpendicular_intersection(A, B, C):
    # Convert coordinates to numpy arrays for easier computation
    A = np.array(A)
    B = np.array(B)
    C = np.array(C)
    
    # Calculate the direction vector of line BC
    BC = C - B
    
    # Compute the slope of BC if not vertical
    if BC[0] != 0:
        slope_BC = BC[1] / BC[0]
        # Slope of the perpendicular line from A to BC
        slope_perpendicular = -1 / slope_BC
    else:
        # If line BC is vertical, then perpendicular line is horizontal
        slope_perpendicular = 0
    
    # Calculate the equation of the line passing through A and perpendicular to BC
    # y - y_A = slope_perpendicular * (x - x_A)
    # Rearrange to standard form Ax + By + C = 0
    if BC[0] != 0:
        A_coeff = -slope_perpendicular
        B_coeff = 1
        C_coeff = -A_coeff * A[0] - B_coeff * A[1]
    else:
        # If BC is vertical, AE must be horizontal
        A_coeff = 1
        B_coeff = 0
        C_coeff = -A[0]
    
    # Equation of line BC: (y - y_B) = slope_BC * (x - x_B)
    # Convert to Ax + By + C = 0 for line intersection calculation
    if BC[0] != 0:
        A_BC = -slope_BC
        B_BC = 1
        C_BC = -A_BC * B[0] - B_BC * B[1]
    else:
        # BC is vertical, so x = constant
        A_BC = 1
        B_BC = 0
        C_BC = -B[0]
    
    # Solve the linear system of equations representing the two lines
    # [A_coeff B_coeff] [x] = [-C_coeff]
    # [A_BC    B_BC   ] [y]   [-C_BC  ]
    matrix = np.array([[A_coeff, B_coeff], [A_BC, B_BC]])
    constants = np.array([-C_coeff, -C_BC])
    
    # Use numpy to solve the linear system
    intersection = np.linalg.solve(matrix, constants)
    return intersection.tolist()


# this function takes a coordinate A, start and end points of a line BC, and returns the coordinates of the point E on BC such that AE is parallel to BC
def find_parallel_intersection(A, B, C):
    # Convert coordinates to numpy arrays for vector operations
    A = np.array(A)
    B = np.array(B)
    C = np.array(C)
    
    # Calculate the direction vector of line BC
    direction_BC = C - B
    
    # Since AE is parallel to BC, its direction vector is the same as BC
    direction_AE = direction_BC
    
    # To find a reasonable "point E", you can just extend AE from A by some length.
    # For visualization, let's extend by the length of BC
    length_BC = np.linalg.norm(direction_BC)
    
    # Normalize the direction vector of AE
    direction_AE_normalized = direction_AE / np.linalg.norm(direction_AE)
    
    # Point E can be found by moving from A in the direction of AE by the length of BC
    E = A + direction_AE_normalized * length_BC
    
    return E.tolist()
'''
# GOAL #: Based on the above tools, I want you to reason about how to draw auxiliary lines on the geometry images provided in matplotlib format to gradually help a student with the problem, then use Code Interpreter to write the code.
# Hint #:  When deciding what type of auxiliary line to draw, consider the following three options: 
(1) Perpendicular line: If you encounter MeasureOf(Angle()),60) or MeasureOf(Angle()),45), you can draw a perpendicular line. Use find_perpendicular_intersection(A, B, C) to find the coordinate of point E such that AE is perpendicular to BC. 
(2) Parallel line: Use find_parallel_intersection(A, B, C) to draw a line from point A parallel to line BC, which returns the coordinate of point E where AE is parallel to BC. Draw Parallel Lines: Parallel lines are always the same distance apart, which helps in geometry by giving equal lengths instantly. It is useful when finding perimeters of a shape. For example, if you draw a line parallel to one side of a shape, you automatically know the opposite side is the same length.
(3) Connecting points in the circle: When there is a circle, connect the circle center to the point in the perimeter
(4) One effective strategy involves using properties of circles, especially the tangent properties and the circle theorem (tangent segments from the same external point to a circle are equal). We can use the given lengths of segments and properties of the circle to form equations. However, for visual aid, drawing auxiliary lines that help visualize relationships or create right triangles might be helpful. Specifically, drawing radii to the points of tangency can aid in identifying and using these properties.

# REQUIREMENTS #:
1. The generated actions can help the student with the user request # USER REQUEST #. You should not give away the answer.
2. You should be concise with your response

Below are some examples of how to use the tools to solve the user requests. You can refer to them for help. You can also refer to the tool descriptions for more information.
# EXAMPLE #:
# USER REQUEST #: Given the geometry diagram <img src='an image'> and the diagram logic form 
            "Equals(LengthOf(Line(Q, R)), 6)",
            "Equals(LengthOf(Line(U, T)), 4)",
            "Equals(LengthOf(Line(V, W)), 8)",
            "Equals(LengthOf(Line(W, P)), 3)",
            "Equals(LengthOf(Line(R, T)), x)",
            "PointLiesOnLine(W, Line(V, P))",
            "PointLiesOnLine(U, Line(V, T))",
            "PointLiesOnLine(Q, Line(P, R))",
            "PointLiesOnLine(S, Line(T, R))",
            "PointLiesOnCircle(U, Circle(C, radius_6_0))",
            "PointLiesOnCircle(S, Circle(C, radius_6_0))",
            "PointLiesOnCircle(W, Circle(C, radius_6_0))",
            "PointLiesOnCircle(Q, Circle(C, radius_6_0))", 
            Below is the original matplotlib code of the geometry: "import matplotlib.pyplot as plt\nimport numpy as np\n\n# Define coordinates\npoints = {\n    \"C\": [153.0, 88.0], \n    \"P\": [114.0, 4.0], \n    \"Q\": [169.0, 4.0], \n    \"R\": [269.0, 5.0], \n    \"S\": [231.0, 116.0], \n    \"T\": [212.0, 171.0], \n    \"U\": [154.0, 171.0], \n    \"V\": [0.0, 173.0], \n    \"W\": [70.0, 69.0]\n}\n\n# Define lines\nlines = [\n    (\"R\", \"Q\"), (\"R\", \"S\"), (\"P\", \"Q\"), (\"P\", \"R\"), (\"P\", \"W\"), \n    (\"T\", \"R\"), (\"T\", \"S\"), (\"U\", \"T\"), (\"V\", \"P\"), (\"V\", \"T\"), \n    (\"V\", \"U\"), (\"V\", \"W\")\n]\n\n# Create plot\nfig, ax = plt.subplots()\nax.set_aspect('equal')\nax.axis('off')\n\n# Draw lines\nfor line in lines:\n    x_values = [points[line[0]][0], points[line[1]][0]]\n    y_values = [points[line[0]][1], points[line[1]][1]]\n    ax.plot(x_values, y_values, 'k')\n\n# Draw circle with radius as the largest distance from C\ncircle_radius = 0\nfor point in ['U', 'S', 'W', 'Q']:\n    dist = np.sqrt((points[point][0] - points[\"C\"][0])**2 + (points[point][1] - points[\"C\"][1])**2)\n    circle_radius = max(circle_radius, dist)\n\ncircle = plt.Circle((points[\"C\"][0], points[\"C\"][1]), circle_radius, color='k', fill=False)\nax.add_artist(circle)\n\n# Set limits\nax.set_xlim(points[\"C\"][0] - 2 * circle_radius, points[\"C\"][0] + 2 * circle_radius)\nax.set_ylim(points[\"C\"][1] - 2 * circle_radius, points[\"C\"][1] + 2 * circle_radius)\n\n# Plot points and labels\nfor point, coord in points.items():\n    ax.plot(coord[0], coord[1], 'ro')\n    ax.text(coord[0], coord[1], f' {point}', fontsize=20, color='red', verticalalignment='bottom', horizontalalignment='left')\n\nplt.show()\n",
            you must draw auxiliary lines to solve the following question: Find x. Assume that segments that appear to be tangent are tangent.\n
# RESULT #:
THOUGHT 0: To solve for x, which is the length of line segment RT, we'll utilize the information provided about the geometric relationships, particularly those involving circles and tangencies. Given that U, S, W, and Q lie on the same circle centered at C with radius 6, and segments that appear tangent to this circle are tangent, we can use these properties to determine x.
ACTION 0: 
Use Code Interpreter to write the code:
import matplotlib.pyplot as plt
import numpy as np

# Define coordinates
points = {
    "C": [153.0, 88.0], 
    "P": [114.0, 4.0], 
    "Q": [169.0, 4.0], 
    "R": [269.0, 5.0], 
    "S": [231.0, 116.0], 
    "T": [212.0, 171.0], 
    "U": [154.0, 171.0], 
    "V": [0.0, 173.0], 
    "W": [70.0, 69.0]
}

# Define lines
lines = [
    ("R", "Q"), ("R", "S"), ("P", "Q"), ("P", "R"), ("P", "W"), 
    ("T", "R"), ("T", "S"), ("U", "T"), ("V", "P"), ("V", "T"), 
    ("V", "U"), ("V", "W")
]

# Create plot
fig, ax = plt.subplots()
ax.set_aspect('equal')
ax.axis('off')

# Draw lines
for line in lines:
    x_values = [points[line[0]][0], points[line[1]][0]]
    y_values = [points[line[0]][1], points[line[1]][1]]
    ax.plot(x_values, y_values, 'k')

# Draw circle with radius as the largest distance from C
circle_radius = 0
for point in ['U', 'S', 'W', 'Q']:
    dist = np.sqrt((points[point][0] - points["C"][0])**2 + (points[point][1] - points["C"][1])**2)
    circle_radius = max(circle_radius, dist)

circle = plt.Circle((points["C"][0], points["C"][1]), circle_radius, color='k', fill=False)
ax.add_artist(circle)

# Draw auxiliary lines (radii to points U, S, W, Q)
auxiliary_lines = [("C", "U"), ("C", "S"), ("C", "W"), ("C", "Q")]
for line in auxiliary_lines:
    x_values = [points[line[0]][0], points[line[1]][0]]
    y_values = [points[line[0]][1], points[line[1]][1]]
    ax.plot(x_values, y_values, 'b--')  # Draw in blue dashed line for clarity

# Set limits
ax.set_xlim(points["C"][0] - 2 * circle_radius, points["C"][0] + 2 * circle_radius)
ax.set_ylim(points["C"][1] - 2 * circle_radius, points["C"][1] + 2 * circle_radius)

# Plot points and labels
for point, coord in points.items():
    ax.plot(coord[0], coord[1], 'ro')
    ax.text(coord[0], coord[1], f' {point}', fontsize=20, color='red', verticalalignment='bottom', horizontalalignment='left')

plt.show()

OBSERVATION: Execution success. The output is as follows:
<the image outputs of the previous code is here.> 
If you can get the answer, please reply with ANSWER: <your answer> and ends with TERMINATE. Otherwise, please generate the next THOUGHT and ACTION.
THOUGHT 1: To solve X, we can ....
ACTION 1: No action needed.
ANSWER: 10. TERMINATE
\end{lstlisting}

\subsection{System Prompt for Direct Problem Solving}
The system prompt used for comparing performance against Visual Sketchpad on the IsoBench dataset.

\begin{lstlisting}
        You are a personal tutor. When asked a question, write and run code to draw a helpful diagram to help solve the question, then use the diagram to solve the question.
\end{lstlisting}
\label{subsec:prompt_solve_problem}

\subsection{Prompt Footer for IsoBench}
\label{prompt_footer}

The following prompt footer was appended to all of the task specific prompts used in IsoBench.
\begin{lstlisting}
First give an explanation of your answer:
Explanation: <explanation>

Then give your final answer as
ANSWER: <answer>

You should give the final answer without formatting and without any additional information.
\end{lstlisting}

\end{document}